\documentclass[useAMS,usenatbib]{mn2e}
\usepackage[english]{babel}

\usepackage{fancyhdr}
\usepackage{amsfonts}
\usepackage{amsmath}
\usepackage{amssymb}
\usepackage{multicol}
\usepackage{layout}
\usepackage{graphicx}
\usepackage{subfigure}

\usepackage{times}
\usepackage{natbib}

\newif\ifAMStwofonts
\AMStwofontstrue

\newcommand{\aj}{AJ}
\newcommand{\apj}{ApJ}
\newcommand{\apjs}{ApJS}
\newcommand{\aap}{A\&A}
\newcommand{\mnras}{MNRAS}
\newcommand{\prd}{Phys. Rev. D}

\graphicspath{{./fig/}}

\title[Spherical collapse model in dark energy cosmologies]
    {Spherical collapse model in dark energy cosmologies}

\author[F. Pace et al.]
       {F. Pace$^{1}$\thanks{E-mail: francesco@ita.uni-heidelberg.de}, 
         J.-C. Waizmann$^{1}$, M. Bartelmann$^{1}$\\
         $^{1}$Institut f\"ur Theoretische Astrophysik, Zentrum f\"ur
         Astronomie, Universit\"at Heidelberg,
         Albert-Ueberle-Stra\ss e 2, 69120 Heidelberg, Germany}

\date{Received \today; accepted ?}

\pagerange{\pageref{firstpage}--\pageref{lastpage}} \pubyear{2009}

\begin{document}
\label{firstpage}
\maketitle

\begin{abstract}
We study the spherical collapse model for several dark energy scenarios using the fully nonlinear differential equation for the evolution of the density contrast within homogeneous spherical overdensities derived from Newtonian hydrodynamics. While mathematically equivalent to the more common approach based on the differential equation for the radius of the perturbation, this approach has substantial conceptual as well as numerical advantages. Among the most important are that no singularities at early times appear, which avoids numerical problems in particular in applications to cosmologies with dynamical and early dark energy, and that the assumption of time-reversal symmetry can easily be dropped where it is not strictly satisfied. We use this approach to derive the two parameters characterising the spherical-collapse model, i.e.~the linear density threshold for collapse $\delta_\mathrm{c}$ and the virial overdensity $\Delta_\mathrm{V}$, for a broad variety of dark-energy models and to reconsider these parameters in cosmologies with early dark energy. We find that, independently of the model under investigation, $\delta_\mathrm{c}$ and $\Delta_\mathrm{V}$ are always very close to the values obtained for the standard $\Lambda$CDM model, arguing that the abundance of and the mean density within non-linear structures are quite insensitive to the differences between dark-energy cosmologies. Regarding early dark energy, we thus arrive at a different conclusion than some earlier papers, including one from our group, and we explain why.
\end{abstract}


\section{Introduction}

Over the last decade, a wealth of evidence was accumulated in favour of the conclusion that the expansion of our Universe is accelerating, mainly from the observation of the type-Ia supernovae \citep{Riess1998, Perlmutter1999, Kowalski2008} and the cosmic microwave background (CMB) \citep{Komatsu2009} in combination with measurements of the Hubble constant and large-scale structures (LSS) \citep{Cole2005}. Assuming the validity of general relativity on large scales, a possible explanation for this accelerating expansion, is obtained by introducing a component of the cosmic fluid, the dark energy, with equation-of-state parameter $w<-1/3$.

Despite efforts from both observational and theoretical sides, the nature of the dark energy remains obscure. Consequently, a plethora of different models has been proposed for the origin and the time evolution of the dark energy, see for example \cite{Copeland2006} for a comprehensive review. The simplest model assumes that the dark energy is connected with the vacuum energy, the so called cosmological constant, with equation-of-state parameter $w=-1$. Despite the fact that observations constrain the present value of $w$ quite tightly, the time evolution of the equation-of-state parameter is rather poorly constrained. Thus, it is natural to study more general classes of models allowing a time evolution of the dark-energy component, such as models involving scalar fields.

Scalar fields occur naturally in particle physics and in string theory and could thus be candidates to explain the nature of the dark-energy if they are sufficiently strongly self-interacting. This class includes quintessence models, phantom models, K-essence, tachyon models and so forth. Scalar fields are described by their Lagrangian density with a kinetic term $\dot{\phi}^2/2$ and a potential term $V(\phi)$. The equation-of-state parameter then follows from the canonical energy-momentum tensor. If the dark energy is spatially homogeneous, $P=\dot{\phi}^2/2+V(\phi)$ and $\rho=\dot{\phi}^2/2-V(\phi)$, giving $w=P/(\rho c^2)$.

The cosmological-constant case is recovered if the kinetic energy is negligibly small compared to the potential energy. Dark energy affects first of all the expansion rate, causing geometrical effects that can be revealed through distance measurements, such as the luminosity distance to distant supernovae. Secondly, it affects structure formation, the early stages of which can be quantified by the growth factor. Thus, structure formation will be affected by the amount of dark energy and by its dynamical evolution over cosmic history.

One model recently suggested as a candidate for solving the fine-tuning problem of the cosmological constant was the class of early dark-energy cosmologies (EDE) \citep{Wetterich2004,Doran2006}, according to which the contribution of dark energy at early times is not negligible. Thus, to produce the same amount of structure now, structure formation should start earlier and proceed more slowly than in the common $\Lambda$CDM model. It would thus compensate the additional, opposing effects of the early dark-energy contribution. An analytic calculation based on the evolution equation for the radius of a spherical, homogeneous perturbation and various assumptions \citep{Bartelmann2006} implied a substantial increase in the number of objects compared to a standard $\Lambda$CDM model. While this expectation was confirmed by \cite{Sadeh2007}, subsequent $N$-body simulations by \cite{Francis2009b} and \cite{Grossi2009} found instead that the effect of EDE on the mass function of dark-matter haloes and its evolution is almost negligible: The EDE class of models predicts differences in the mass function of only a few percent with respect to the $\Lambda$CDM model. A new numerical derivation of the linear overdensity parameter, also based on the differential equations of the spherical collapse model \citep{Francis2009a} was in perfect agreement with the numerical simulations.

Motivated in part by this discrepancy of results derived from the same model, we are here addressing the problem of determining the time evolution of the linear overdensity $\delta_c(z)$ in a completely different way, using a perturbative approach based on Newtonian hydrodynamics directly. One of the advantages, apart from increased numerical stability, is that no time-reversal symmetry needs to be assumed for the spherical collapse.

The structure of the paper is as follows. In Sect.~\ref{sect:nhydro}, we present the basic equations of Newtonian hydrodynamics and sketch the derivation of the equations used to obtain the linear over density threshold $\delta_{\mathrm{c}}$ and the non-linear virial overdensity parameter $\Delta_{\mathrm{V}}$. In Sect.~\ref{sect:models}, we briefly describe and motivate the cosmological models investigated in this work, while we compare them in Sect.~\ref{sect:res} with the $\Lambda$CDM model. We present our conclusions in Sect.~\ref{sect:conc}. In the appendix \ref{sect:app}, we discuss why the previous theoretical estimations of $\delta_\mathrm{c}$ for the EDE models obtained by \cite{Bartelmann2006} and \cite{Sadeh2007} differ from the results obtained in this work and by \cite{Francis2009a}.

\section{Newtonian hydrodynamics of a relativistic fluid}\label{sect:nhydro}

We review here the derivation of the differential equation determining the evolution of an overdensity $\delta$. The final non-linear equation specialised to $w=0$ is not new, but has already been used by several authors in the context of structure formation \citep{Padmanabhan1996,Abramo2007} and for the study of the spherical and ellipsoidal collapse \citep{Bernardeau1994, Ohta2003, Ohta2004}. The linearised equation was presented in \cite{ColesLucchin2002} specialised for two limiting cases, namely dust ($w=0$) and relativistic matter ($w=1/3$), and in \cite{Lima1997} for a general model with constant $w$.

Our study, based on the work by \cite{Abramo2007} where the equation for the evolution of the overdensity $\delta$ was generalised to allow for time-dependent equation-of-state for the dark energy component, has two novel aspects: First, we generalise the evolution equation to a generic collapse geometry, rendering the spherical and ellipsoidal models special cases. Second, the newly obtained generality of the method allows its application to modified-gravity cosmologies and coupled-quintessence models. This will be postponed to future work.

Following the work by \cite{Abramo2007}, we derive our equation including the pressure terms explicitly and assuming that the fluid satisfies the equation-of-state $P=w\rho c^2$. Starting from this point, the final equation will be in its most general form and can then be specified to a particular fluid simply by adopting the appropriate equation-of-state.

We start from the continuity equation for the energy-momentum tensor in general relativity, $\nabla_{\nu}T^{\mu\nu}=0$. For a perfect fluid, we have
\begin{equation}
  T^{\mu\nu}=(\rho c^2+P)u^{\mu}u^{\nu}+Pg^{\mu\nu}\;,
\end{equation}
where $\rho$ is the density of the fluid, $P$ its pressure, $u$ its 4-velocity and $g^{\mu\nu}$ the metric.

Contracting the continuity equation once with $u_{\mu}$ and once with the projection operator $g_{\mu\alpha}+u_{\mu}u_{\alpha}$ one obtains the relativistic expressions for the continuity and the Euler equations, respectively:
\begin{eqnarray}
  \frac{\partial\rho}{\partial t}+\nabla_{\vec{r}}\cdot(\rho\vec{v})+
  \frac{P}{c^2}\nabla_{\vec{r}}\cdot\vec{v} = 0 \label{eqn:cnpert}\;,\\
  \frac{\partial\vec{v}}{\partial
    t}+(\vec{v}\cdot\nabla_{\vec{r}})\vec{v}+
  \nabla_{\vec{r}}\Phi+\frac{c^2\nabla_{\vec{r}}P+\vec{v}\dot P}{\rho c^2+P}=0\;. \label{eqn:enpert}
\end{eqnarray}
Here $\vec v$ is the velocity in three-space, $\Phi$ is the Newtonian gravitational potential and $\vec{r}$ is the physical coordinate.

The $0$-$0$ component of Einstein's field equations gives the relativistic Poisson equation
\begin{equation}\label{eqn:pnpert}
  \nabla^2\Phi=4\pi G\left(\rho+\frac{3P}{c^2}\right)\;.
\end{equation}
The continuity equation for the mean background density, obtained from the spatial components of Einstein's equations, is now modified to
\begin{equation}
 \dot{\bar{\rho}}+3H\left(\bar{\rho}+\frac{P}{c^2}\right)=0\;,
\end{equation}
where $\bar{\rho}=\frac{3H^2\Omega_{\mathrm{fluid}}}{8\pi G}$ is the background mass density of all contributions to the cosmic fluid, and $\Omega_{\mathrm{fluid}}$ is its density parameter.

As usual, we introduce comoving coordinates $\vec{x}=\vec{r}/a$ and define
\begin{eqnarray}
 \rho(\vec{x},t) & = & \bar{\rho}(1+\delta(\vec{x},t))\;, \label{eqn:rpert} \\
 P(\vec{x},t) & = & w\rho(\vec{x},t) c^2\;, \label{eqn:ppert}\\
 \Phi(\vec{x},t) & = & \Phi_0(\vec{x},t)+\phi(\vec{x},t)\;, \label{eqn:fpert}\\
 \vec{v}(\vec{x},t) & = & a[H(a)\vec{x}+\vec{u}(\vec{x},t)]\;, \label{eqn:vpert}
\end{eqnarray}
where $H(a)$ is the Hubble function and $\vec{u}(\vec{x},t)$ is the comoving peculiar velocity. Inserting Eqs.~(\ref{eqn:rpert})--(\ref{eqn:vpert}) into Eqs.~(\ref{eqn:cnpert})--(\ref{eqn:pnpert}), we find the equations

\begin{eqnarray}
 \dot{\delta}+(1+w)(1+\delta)\nabla_{\vec{x}}\cdot\vec{u} & = & 0\;, \label{eq:pertCont}\\
 \frac{\partial \vec{u}}{\partial t}+2H\vec{u}+(\vec{u}\cdot\nabla_{\vec{x}})\vec{u}+\frac{1}{a^2}\nabla_{\vec{x}}\phi & = & 0\;,\label{eq:pertEuler} \\
 \nabla_{\vec{x}}^2\phi-4\pi G(1+3w)a^2\bar{\rho}\delta & = & 0\;. \label{eq:pertPois}
\end{eqnarray}

We now take the divergence of the Euler equation (\ref{eq:pertEuler}) and recall the decomposition
\begin{equation}
 \nabla\cdot[(\vec{u}\cdot\nabla)\vec{u}] = \frac{1}{3}\theta^2 + \sigma^2 - \omega^2\;, \label{eq:decomposition}
\end{equation}
into the expansion $\theta=\nabla_{\vec{x}}\cdot\vec{u}$, the shear tensor $\sigma^2=\sigma_{ij}\sigma^{ij}$ and the rotation tensor $\omega^2=\omega_{ij}\omega^{ij}$. Next, taking the time derivative of the continuity equation (\ref{eq:pertCont}) and combining all three equations, we arrive at the fully non-linear evolution equation
\begin{equation}\label{eqn:nleq}
 \begin{split}
  \ddot{\delta}+\left(2H-\frac{\dot{w}}{1+w}\right)\dot{\delta}-\frac{4+3w}{3(1+w)}\frac{\dot{\delta}^2}{1+\delta}-&\\
  4\pi G\bar{\rho}(1+w)(1+3w)\delta(1+\delta)-\\
  (1+w)(1+\delta)(\sigma^2-\omega^2) & = 0\;.
 \end{split}
\end{equation}
Note that the shear is a symmetric traceless tensor, while the rotation is antisymmetric. They read
\begin{eqnarray}
 \sigma_{ij} & = & \frac{1}{2}\left(\frac{\partial u^j}{\partial x^i}+\frac{\partial u^i}{\partial x^j}\right)
 -\frac{1}{3}\theta\delta_{ij}\;, \\
 \omega_{ij} & = & \frac{1}{2}\left(\frac{\partial u^j}{\partial x^i}-\frac{\partial u^i}{\partial x^j}\right)\;.
\end{eqnarray}
Specialising Eq.~(\ref{eqn:nleq}) for dust ($w=0$), we recover Eq.~(41) of \cite{Ohta2003}. We notice that Eq.~(\ref{eqn:nleq}) generalises Eq.~(7) of \cite{Abramo2007} to the case of a non-spherical configuration of a rotating fluid.

Changing the independent variable from the time $t$ to the scale factor $a$ using the relation $\partial_t=aH(a)\partial_a$, the evolution equation assumes the form
\begin{equation}\label{eqn:wnldeq}
 \begin{split}
  \delta^{\prime\prime}+\left(\frac{3}{a}+\frac{E^\prime}{E}-\frac{w^\prime}{1+w}\right)\delta^\prime-\frac{4+3w}{3(1+w)}\frac{\delta^{\prime 2}}{1+\delta}-&\\
  \frac{3}{2}\frac{\Omega_{\mathrm{fluid},0}}{a^2E^2(a)}g(a)(1+w)(1+3w)\delta(1+\delta)-&\\
  \frac{1}{aH^2(a)}(1+w)(1+\delta)(\sigma^2-\omega^2)&=0\;,
 \end{split}
\end{equation}
where $\Omega_{\mathrm{fluid},0}$ is the density parameter of the fluid at $a=1$, and $g(a)$ is a function specifying the time evolution of the dark-energy model considered.

In the following, since we are interested in the collapse of a homogeneous sphere, we ignore the rotation and the shear tensors. The shear tensor vanishes for a sphere anyway. We will also restrict the treatment to spherical perturbation filled with dust, having $w=0$ and $g(a)=a^{-3}$. Thus, the non-linear equation to be solved reads
\begin{equation}\label{eqn:nldeq}
 \delta^{\prime\prime}+\left(\frac{3}{a}+\frac{E^\prime}{E}\right)\delta^\prime-\frac{4}{3}\frac{\delta^{\prime 2}}{1+\delta}-
 \frac{3}{2}\frac{\Omega_{\mathrm{m},0}}{a^5E^2(a)}\delta(1+\delta)=0\;.
\end{equation}

We notice that Eq.~(\ref{eqn:wnldeq}) has a singularity when $w=-1$. To see what happens for the cosmological constant, we multiply both sides with $1+w$ and then specialise to $w=-1$. We obtain $\delta'^2/(1+\delta)=0$, implying $\delta=\mathrm{const}$, and with appropriate initial conditions, the constant can be set to zero, so the cosmological constant can not clump as expected.

\subsection{Determination of $\delta_{\mathrm{c}}$ and $\Delta_{\mathrm{V}}$}

The linearised Eq.~(\ref{eqn:nldeq}) reads
\begin{equation}\label{eqn:ldeq}
 \delta^{\prime\prime}+\left(\frac{3}{a}+\frac{E^\prime}{E}\right)\delta^\prime-\frac{3}{2}\frac{\Omega_{\mathrm{m},0}}{a^5E^2}\delta=0\;,
\end{equation}
and its solution, for appropriate initial conditions, will give the \textit{linear} overdensity parameter $\delta$ at any point in time. This equation is also used to determine the time evolution of the growth factor if suitable initial conditions are used.

To determine the appropriate initial conditions, we start by considering Eq.~(\ref{eqn:nldeq}). We know that, since this represents the non-linear evolution of the density contrast, its value at a some chosen collapse time diverges, $\delta\to\infty$. Thus, we search for an initial density contrast such that the $\delta$ solving the non-linear equation diverges at the chosen collapse time. Numerically, we assume this to be achieved once $\delta\ge10^7$. Since the curve $\delta(a)$ representing the non-linear density evolution turns very steep towards the collapse, the result is very insensitive to the exact choice of this threshold value as long as it is a large number. Once the initial overdensity is found, we use this value as an initial condition in Eq.~(\ref{eqn:ldeq}) to find $\delta_\mathrm{c}$.

Since we are dealing with second-order equations, two initial values have to be given, one for the initial overdensity $\delta_i$ and the other for the initial rate of evolution, $\delta_\mathrm{i}^\prime$. We know that initially, $\delta_\mathrm{i}^\prime$ for the sphere should be small, thus we set it to $\delta_\mathrm{i}^\prime=5\times 10^{-5}$, corresponding to the initial scale factor used for starting the integration of the two differential equations. We carried out several numerical tests to check the dependence of the solution on $\delta_\mathrm{i}^\prime$ and found that the result does not depend on the precise value of $\delta_\mathrm{i}'$. Specifically, we considered several values in the interval between $10^{-6}$ and $10^{-4}$ and saw perfect convergence of the solution. Also setting $\delta_\mathrm{i}'=0$, the result does not change considerably.

\begin{figure}
\includegraphics[angle=-90,width=\hsize]{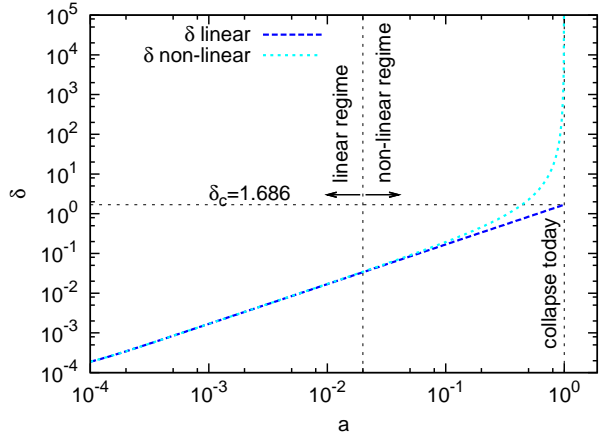}
\caption{Linear (blue dashed curve) and non-linear (cyan short-dashed curve) evolution of the overdensity parameter $\delta$. An EdS model and a sphere collapsing at $z=0$ are assumed. We notice how, after the initial parallel evolution, the non-linear solution grows very fast with the scale factor, in comparison to the linear solution.}
\label{fig:evol}
\end{figure}

In Fig.~\ref{fig:evol}, we show the solution $\delta$ of the non-linear (cyan short-dashed curve) and the linear (blue dashed curve) evolution equations as a function of the scale factor, for an EdS model, supposing that the sphere collapses at $z=0$. We see that the linear solution grows linearly with time, reaching the correct value for $\delta_\mathrm{c}=1.686$ at the collapse scale factor, while, after developing in parallel initially, the non-linear solution starts deviating and growing exponentially. To obtain the virial overdensity, supposing that dark energy does not clump, we follow the prescription of \cite{Maor2005}, which generalises the work by \cite{Wang1998}.

Knowing the non-linear time evolution, it is possible to infer all the other properties of the collapsing sphere, in particular the time evolution of the radius, the turn-around scale factor $a_{\mathrm{ta}}$ when the sphere reaches its maximum radius, and the overdensity at turn-around $\zeta$. The virial overdensity is defined as $\Delta_{\mathrm{V}}=\delta_{\mathrm{nl}}+1=\zeta(x/y)^3$, where $x=a/a_{\mathrm{ta}}$ is the normalised scale factor and $y$ is the radius of the sphere normalised to its value at the turn-around.

To determine the turn-around scale factor, we solve Eq.~(\ref{eqn:wnldeq}) and determine the quantity $\log(\delta_{\mathrm{nl}}+1)/a^3$. Apart from a multiplicative constant, this is the inverse of the collapsing sphere's radius and assumes a minimum at the turn-around scale factor $a_{\mathrm{ta}}$. To determine the virial overdensity at turn-around $\zeta$, we integrate Eq.~(\ref{eqn:wnldeq}) up to $a_{\mathrm{ta}}$ and add the result to unity.

\section{The models}\label{sect:models}

As mentioned before, the nature of the dark energy is still unknown, which leaves room for a plethora of phenomenological or \textit{ad hoc} models being discussed in the literature. Here, we briefly review some models, characterised by the requirements that they try to explain the accelerated expansion of the Universe in terms of a smooth and slowly varying component, the dark energy, quantified by a certain equation-of-state parameter $w(a)$ and formulated in the framework of general relativity. The dark-energy component satisfies the following evolution equation
\begin{equation}\label{eq:deceq}
  \dot{\rho}+3H\left(\rho+\frac{P}{c^2}\right)=\dot{\rho}+3H(1+w)\rho=0\;.
\end{equation}
Our collection of models is based on the works by \cite{Szydlowski2006} and \cite{Jennings2009} where $N$-body simulations of different quintessence models are studied. We shall use the following cosmological parameters: $\Omega_{m,0}=0.274$, $\Omega_{\mathrm{Q},0}=0.726$. For flat models $\Omega_{K,0}=0$, while for models with a curvature term we set $\Omega_{K,0}=-0.018$.

\subsection{$\Lambda$CDM model}

The simplest model used to fit the data that explains the late-time accelerated expansion of the Universe has a cosmological constant, with equation-of-state parameter $w=-1$ independent of time. Because of this, the contribution of the cosmological constant starts dominating only recently and becomes rapidly negligible towards higher redshift, such that, at high redshift, it converges towards the EdS model. Despite conceptual problems associated with it, it is currently the simplest model fitting virtually all available observational data. We shall thus assume a spatially flat $\Lambda$CDM model as a reference model, since observations suggest negligible spatial curvature. However, we will also consider finite curvature, even if this parameter, according to the current limits, is quite small.

\subsection{Quintessence models}

An immediate extension of the cosmological-constant scenario is described by a scalar field very weakly or not interacting with the matter component. This scalar field can be in principle the inflaton itself, even if the vast majority of the scenarios assumes it to be independent of the scalar field actually driving the observed accelerated expansion. These models are described by a kinetic energy and a potential energy characterised by a given functional form, that can either be motivated by theory or introduced \textit{ad hoc}, such as power-law potentials. The equivalent mass of the scalar field is given by the second derivative of the potential term. Compared to the cosmological-constant case, these models have a time-evolving equation-of-state parameter. They are justified by the fact that a time-dependent equation-of-state parameter, not excluded by observations, naturally arises in the framework of a scalar field theory. It is assumed that dark energy does not clump, at least not on the relevant scales accessible to cosmological studies. The sound speed of quintessence models is directly related to the equation-of-state parameter by the relation $c_{\mathrm{s}}=\sqrt{w}c$.

Here, together with the early dark-energy models (see Sect.~\ref{sect:ede}), we study the six models used by \cite{Jennings2009} to which we refer the reader for more detail. All models can be divided into two broad classes, tracking scalar fields \citep{Steinhardt1999} and scaling fields \citep{Halliwell1987, Wands1993, Wetterich1995}. Tracker fields are characterised by a scalar field rolling down its potential reaching an attractor solution. An interesting feature is that the scalar field tracks the dominant component of the cosmic fluid. The second class instead keeps the ratio between the density of the scalar field and that of the background constant. The models INV1 and INV2 have inverse power-law potentials \citep{Corasaniti2003, Sanchez2009, Corasaniti2004}, the SUGRA model is a typical example for tracking behaviour as well as the CNR model \citep{Copeland2000}, while the models 2EXP \citep{Barreiro2000} and AS \citep{Albrecht2000} are examples for scaling fields.

Given the appropriate equation-of-state parameter for each model, we can solve the continuity equation \ref{eq:deceq}, leading to the solution
\begin{equation}
  \rho=\rho_0 e^{-3\int_1^a [1+w(a^\prime)]d\ln a^\prime} \;.
\end{equation}
Thus, the expansion function reads
\begin{equation}\label{eq:e}
  E(a)=\sqrt{\frac{\Omega_{\mathrm{m},0}}{a^3}+
  \frac{\Omega_{\mathrm{K},0}}{a^2}+
  \Omega_{\mathrm{Q},0}\exp{\left(-3\int_1^a \frac{1+w(a')}{a'}da'\right)}}\;.
\end{equation}
All these models can be described by the following equation-of-state parameter, valid after matter-radiation equality,
\begin{equation}
w(a)=w_0+(w_{\mathrm{m}}-w_0)\frac{1+e^{\frac{a_{\mathrm{m}}}{\Delta_{\mathrm{m}}}}}{1+e^{-\frac{a-a_{\mathrm{m}}}{\Delta_{\mathrm{m}}}}}\frac{1-e^{-\frac{a-1}{\Delta_{\mathrm{m}}}}}{1-e^{\frac{1}{\Delta_{\mathrm{m}}}}}\;.
\end{equation}

In Tab.~\ref{tab:param} we list the values of the parameters $a_{\mathrm{m}}$, $\Delta_{\mathrm{m}}$, $w_{\mathrm{m}}$ and $w_0$ characterising the models discussed above.
\begin{table}\label{tab:param}
\caption{Parameter values for the quintessence models.}
\begin{center}
\begin{tabular}{c|c|c|c|c|c|}
\hline
\hline
Model & $w_0$ & $w_{\mathrm{m}}$ & $a_{\mathrm{m}}$ & $\Delta_{\mathrm{m}}$ \\
\hline
INV1  & -0.4  & -0.27 & 0.18  & 0.5        \\
INV2  & -0.79 & -0.67 & 0.29  & 0.4        \\
2EXP  & -1.0  &  0.01 & 0.19  & 0.043      \\
AS    & -0.96 & -0.01 & 0.53  & 0.13       \\
CNR   & -1.0  &  0.1  & 0.15  & 0.016      \\
SUGRA & -0.82 & -0.18 & 0.1   & 0.7        \\
\hline
\end{tabular}
\end{center}
\end{table}

The CPL model \cite{Chevallier2001,Linder2003} has the equation-of-state parameter
\begin{equation}
  w(z)=-1+\frac{z}{1+z} \;,
\end{equation}
mimicking a cosmological constant at low redshift and growing up to zero at very early times.

\begin{figure}
\includegraphics[angle=-90,width=\hsize]{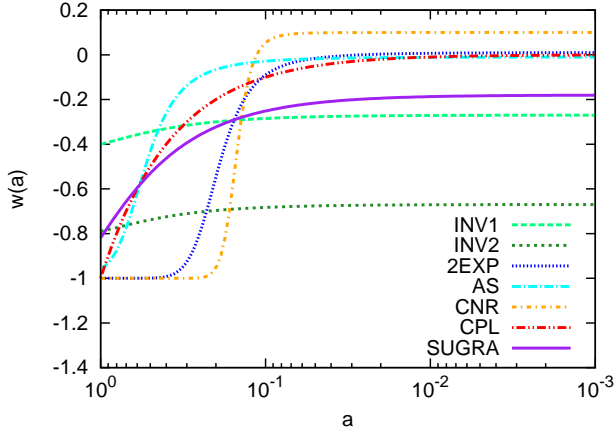}
\caption{Time evolution of the equation-of-state parameter as a function of the scale factor $a$ for the different quintessence models studied in this work. The light-green dashed and the dark-green short-dashed curves represent the INV1 and INV2 models, respectively; the blue dotted curve the 2EXP model, the cyan dot-dashed curve the AS model, the orange dot-short-dashed and red dot-dot-dashed curves the CNR and the CPL models, while the solid violet curve represents the SUGRA model.}
\label{fig:quint}
\end{figure}

In Fig.~\ref{fig:quint} we show the equation-of-state parameter of as a function of the scale factor $a$ for the different models. The linestyle coding is given in the figure. We notice that these models show different behaviour: the INV1 and INV2 models show a gentle increase of the equation-of-state parameter, which is almost constant, except for late times. The other models instead show a very large change in $w$ at late times, reaching a constant value quite soon in cosmic history. Also, the values of the equation-of-state parameter at $a=1$ cover a broad range, from $w=-0.4$ for the INV1 model till $w=-1$ for essentially all others.

Even if the SUGRA model was studied both analytically and numerically by \cite{Mainini2003b} and \cite{Mainini2003a}, we determine once again the expected $\delta_{\mathrm{c}}$ for this model as a test of the validity of our approach and for completeness.

\subsubsection{Early dark energy models}\label{sect:ede}

A particular class of dark-energy models, introduced by \cite{Wetterich2004} and studied in detail by \cite{Doran2006}, introducing a convenient functional form for its equation-of-state parameter, has a small but finite density of early dark energy (EDE). This class of model, where the density parameter of the dark-energy component remains at the level of a few percent at very early times, was used by \cite{Bartelmann2006} for their study of non-linear structure formation.

Its equation-of-state parameter is implicitly given by
\begin{equation}
  \left[3w(a)-\frac{a_{\mathrm{eq}}}{a+a_{\mathrm{eq}}}\right]
  \Omega_{\mathrm{Q}}(a)(1-\Omega_{\mathrm{Q}}(a))= -\frac{d\Omega_{\mathrm{Q}}(a)}{d\ln a}\;,
\end{equation}
where $a_{\mathrm{eq}}$ is the scale factor at matter-radiation equality, and $\Omega_{\mathrm{Q}}(a)$ represents the time evolution of the dark energy component,
\begin{equation}
  \Omega_{\mathrm{Q}}(a)=\frac{\Omega_{\mathrm{Q},0}-\Omega_{\mathrm{Q},\mathrm{e}}(1-a^{-3w_0})^\gamma}
  {\Omega_{\mathrm{Q},0}+ \Omega_{\mathrm{m},0}^{3w_0}}+\Omega_{\mathrm{Q},\mathrm{e}}
  (1-a^{-3w_0})^\gamma\;,
\end{equation}
where $\Omega_{\mathrm{Q},0}$ is the density parameter of dark energy today, $\Omega_{\mathrm{Q},\mathrm{e}}$ its density parameter at early times, $w_0$ the present equation-of-state parameter, and $\gamma$ is a shape parameter controlling the importance of the terms containing $\Omega_{\mathrm{Q},\mathrm{e}}$. The expression explicitly assumes a flat universe. We adopt $\gamma=1$ here. Since the equation-of-state parameter for the EDE model is very similar to that of the quintessence models mentioned above, we do not report it here. Further detail on the comparison of the new approach with the old one to determine $\delta_\mathrm{c}$ for the EDE models will be given in Appendix~\ref{sect:app}, where we also compare our prediction with the numerical mass function by \cite{Grossi2009}.

Here, we use the cosmological parameters of the model EDE4 from \cite{Waizmann2009}, i.e.~$\Omega_{\mathrm{m},0}=0.282$, $\Omega_{\mathrm{Q},0}=0.718$ and $w_0=-0.934$, while for the comparison with the numerical simulations, we use $\Omega_{\mathrm{m},0}=0.25$, $\Omega_{\mathrm{Q},0}=0.75$ and $w_0=-0.99$.

\subsection{Chaplygin gas and Casimir effect}

An alternative to scalar fields for explaining the current acceleration proceeds by specifying an exotic equation of state satisfying the condition for acceleration, $w<-1/3$. One example is the Chaplygin gas, first proposed in aerodynamics and subsequently derived from the action in string theory \citep{Ogawa2000}. It was used in cosmology as a possible alternative to dark-energy models by \cite{Kamenshchik2001, Fabris2002, Szydlowski2004}. The equation of state of the generalised Chaplygin gas assumes the form
\begin{equation}
  P=-\frac{A}{\rho^\alpha}\;,
\end{equation}
where $A>0$ and $\alpha$ are constants. The classical Chaplygin gas is recovered for $\alpha=1$. Using the continuity equation for the generalised Chaplygin gas, one obtains the dependence of the density on the scale factor
\begin{equation}
  \rho=\left[A+\frac{B}{a^{3(\alpha+1)}}\right]^{1/(1+\alpha)}\;,
\end{equation}
where $B$ is an integration constant. The equation-of-state parameter can be written in the form
\begin{equation}
  w(a)=-\frac{A}{A+\frac{B}{a^{3(\alpha+1)}}}
\end{equation}
where $A=-w_0(\Omega_{\mathrm{Q},0}\rho_{\mathrm{c}})^{1+\alpha}$ and $B=(1+w_0)(\Omega_{\mathrm{Q},0}\rho_{\mathrm{c}})^{1+\alpha}$. $\rho_{\mathrm{c}}$ is the present critical density and $w_0=-A/(A+B)$ is the present value of the equation-of-state parameter. We study the classical Chaplygin gas with $\alpha=1$ and a generalised version with $\alpha=0.2$. Both models have $w_0=-0.8$.

Another possible way to explain the current accelerated expansion is to study quantum properties of the vacuum using the Casimir effect. This effect arises from a change in the zero-point oscillation spectrum of a quantised field when the quantisation domain is finite or the space topology is non-trivial. In a cosmological context, the Casimir effect is relevant if the topology is not simply connected or when compact extra dimensions are involved. In a more general setting, it can be used to study the properties of the vacuum. In this context the contribution given by the Casimir force is scaling like $a^{-4}$, exactly like relativistic species do.

The expansion function is
\begin{equation}\label{eq:eC}
  E(a)=\sqrt{\frac{\Omega_{\mathrm{m},0}}{a^3}+\Omega_{\mathrm{Q},0}-\frac{\Omega_{\mathrm{Cass},0}}{a^4}}\;,
\end{equation}
where $\Omega_{\mathrm{Cass},0}$ is the density of the Casimir component today. If one wants to interpret this as a time evolution of the dark energy component, one can invert Eq.~(\ref{eq:eC}) using the general equation
\begin{equation}
  w(a)=-\frac{1+\frac{2}{3}a\frac{d\ln E(a)}{da}-
  \frac{1}{3}\frac{\Omega_{\mathrm{K},0}}{a^2E(a)^2}}{1-\frac{\Omega_{\mathrm{m},0}}{a^3E(a)^2}- 
  \frac{\Omega_{\mathrm{K},0}}{a^2E(a)^2}}
\end{equation}
to obtain the equation-of-state parameter
\begin{equation}
  w(a)=-\frac{1}{3}\frac{3\Omega_{\mathrm{Q},0}a^4+\Omega_{\mathrm{Cass},0}}
  {\Omega_{\mathrm{Q},0}a^4-\Omega_{\mathrm{Cass},0}}\;.
\end{equation}
Here, we shall assume $\Omega_{\mathrm{Cass},0}=-0.00035$.

\begin{figure}
\includegraphics[angle=-90,width=\hsize]{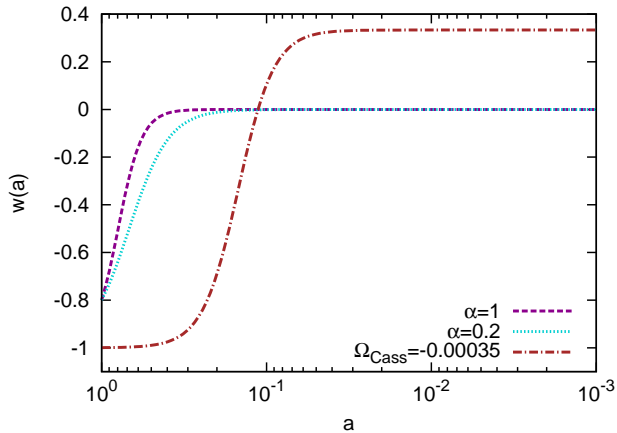}
\caption{Time evolution of the equation-of-state parameter. The (turquoise dotted) magenta dashed curve shows the (generalised) Chaplygin gas while the brown dot-dashed curve represents the model based on the Casimir effect.}
\label{fig:CasChapl}
\end{figure}

In Fig.~\ref{fig:CasChapl}, we show the time evolution of the Chaplygin (magenta dashed curve) and of the generalised Chaplygin (turquoise dotted curve) gas, and in brown dot-dashed the equation-of-state parameter for the Casimir effect. The two curves representing the generalised Chaplygin gas are very similar, only the initial slope changes with the change of the $\alpha$ parameter.

\subsection{Phantom models and topological defects}

A better fit to type-Ia supernova data is achieved if one assumes a varying equation-of-state parameter with some phantom crossing at low redshift, or in other words that there $w<-1$. Models fulfilling this condition are called phantom models, and they seriously challenge the foundations of theoretical physics, since they violate several energy conditions. Here, we study five different phantom models, all having a constant equation-of-state parameter, in particular we focus on the models with $w=-4/3$, $w=-3/2$, $w=-2$, $w=-3$. Certain grand unified theories predict topological defects to have formed in the early universe and since they rapidly diluted, their abundance should be very low. Here, we consider models with $w=-2/3$.

\section{Results for $\delta_{\mathrm{c}}$ and $\Delta_{\mathrm{V}}$}\label{sect:res}

In this section, we discuss the results for the linear overdensity parameter and the virial overdensity for the models introduced in Sect.~\ref{sect:models}, keeping the $\Lambda$CDM model as a reference.

\begin{figure*}
\includegraphics[angle=-90,width=0.49\hsize]{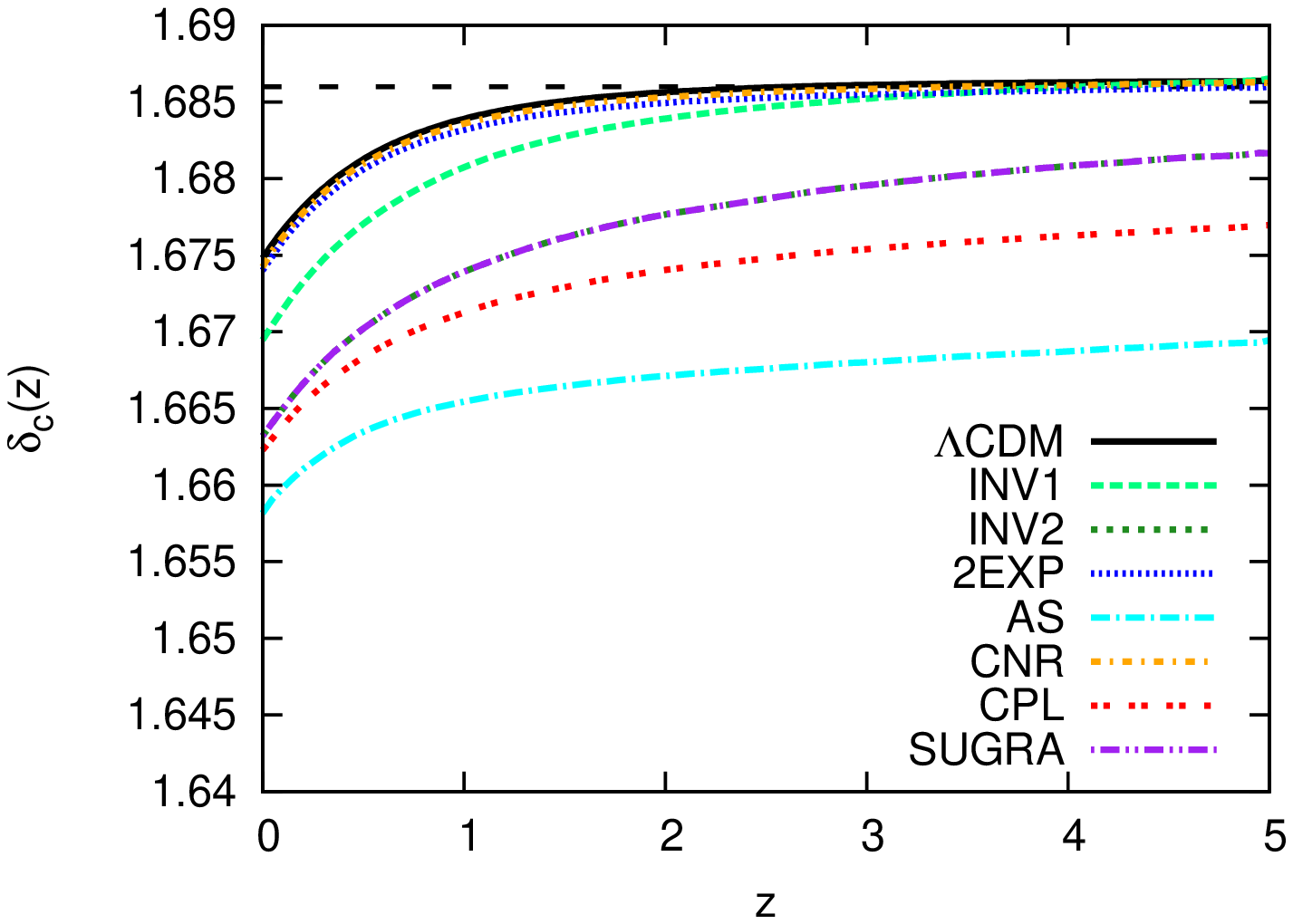}
\includegraphics[angle=-90,width=0.49\hsize]{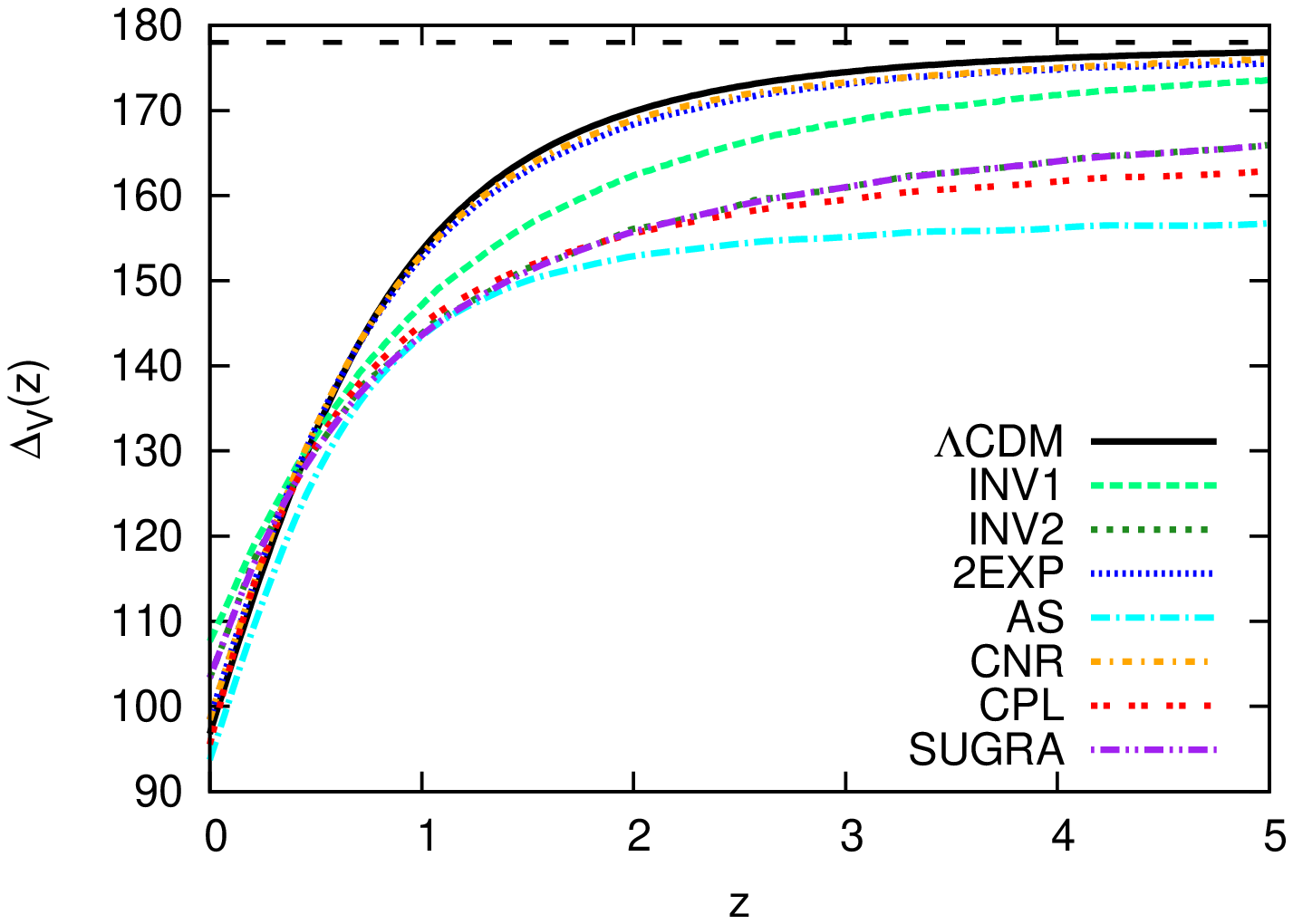}\hfill
\includegraphics[angle=-90,width=0.49\hsize]{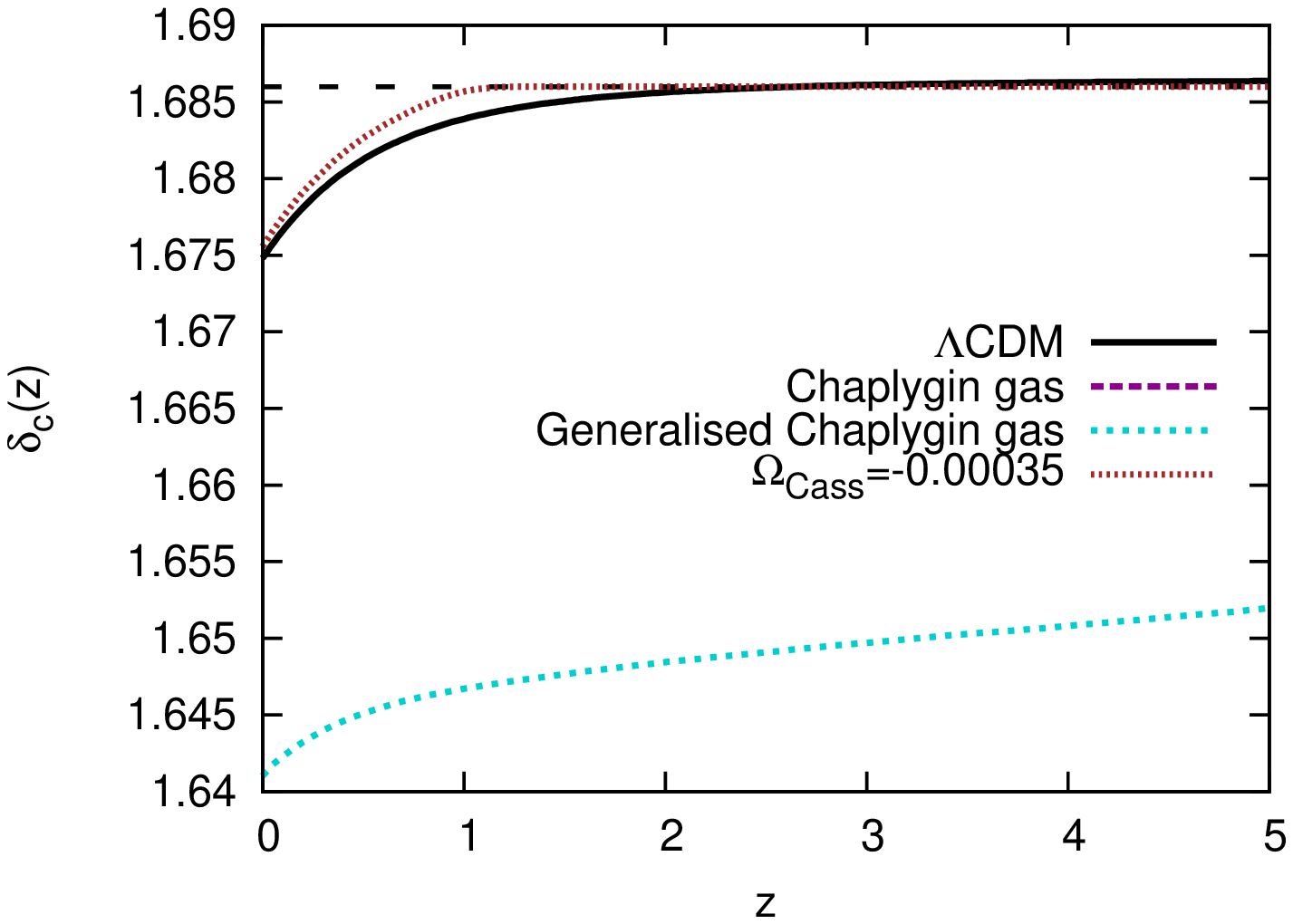}
\includegraphics[angle=-90,width=0.49\hsize]{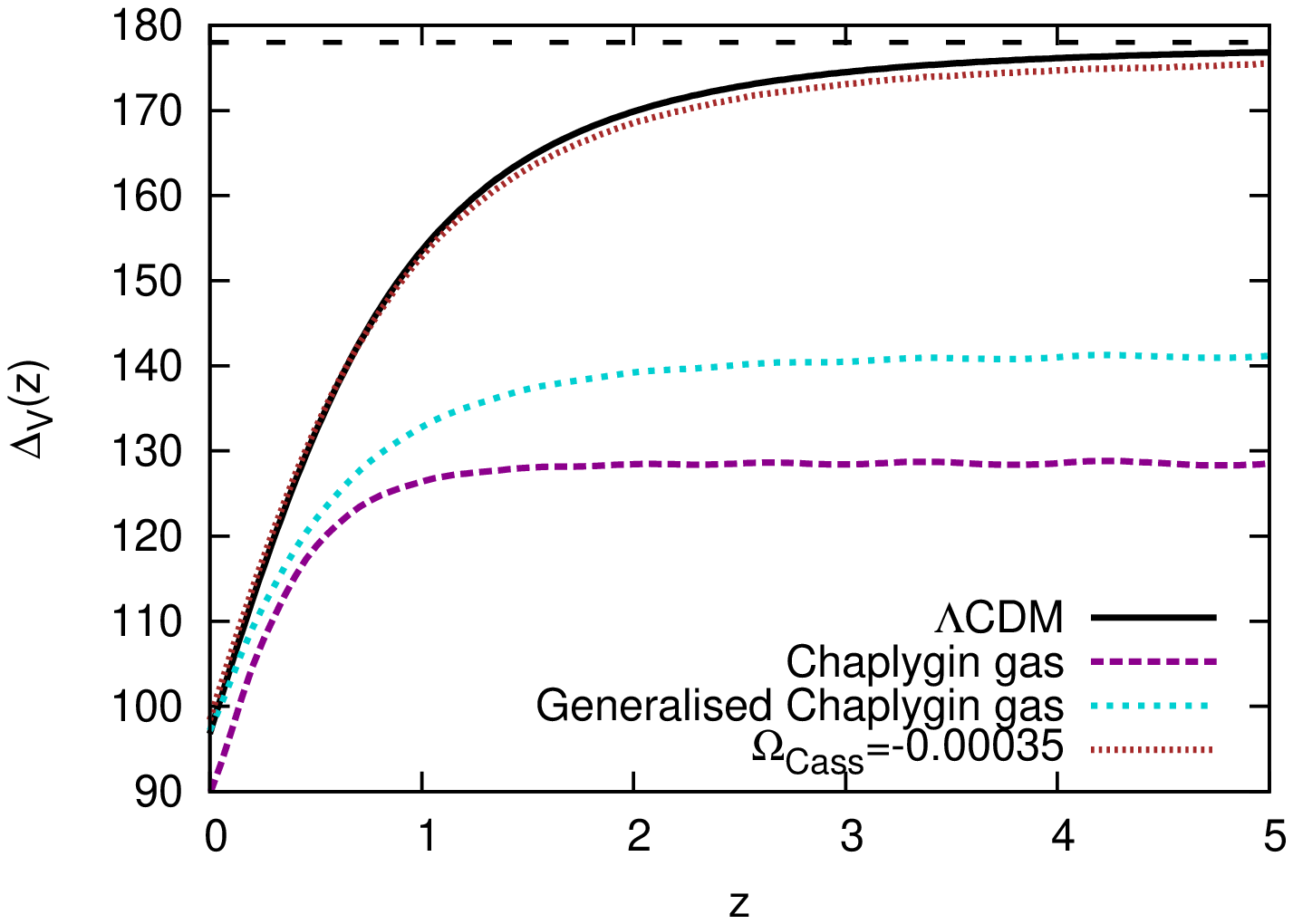}\hfill
\includegraphics[angle=-90,width=0.49\hsize]{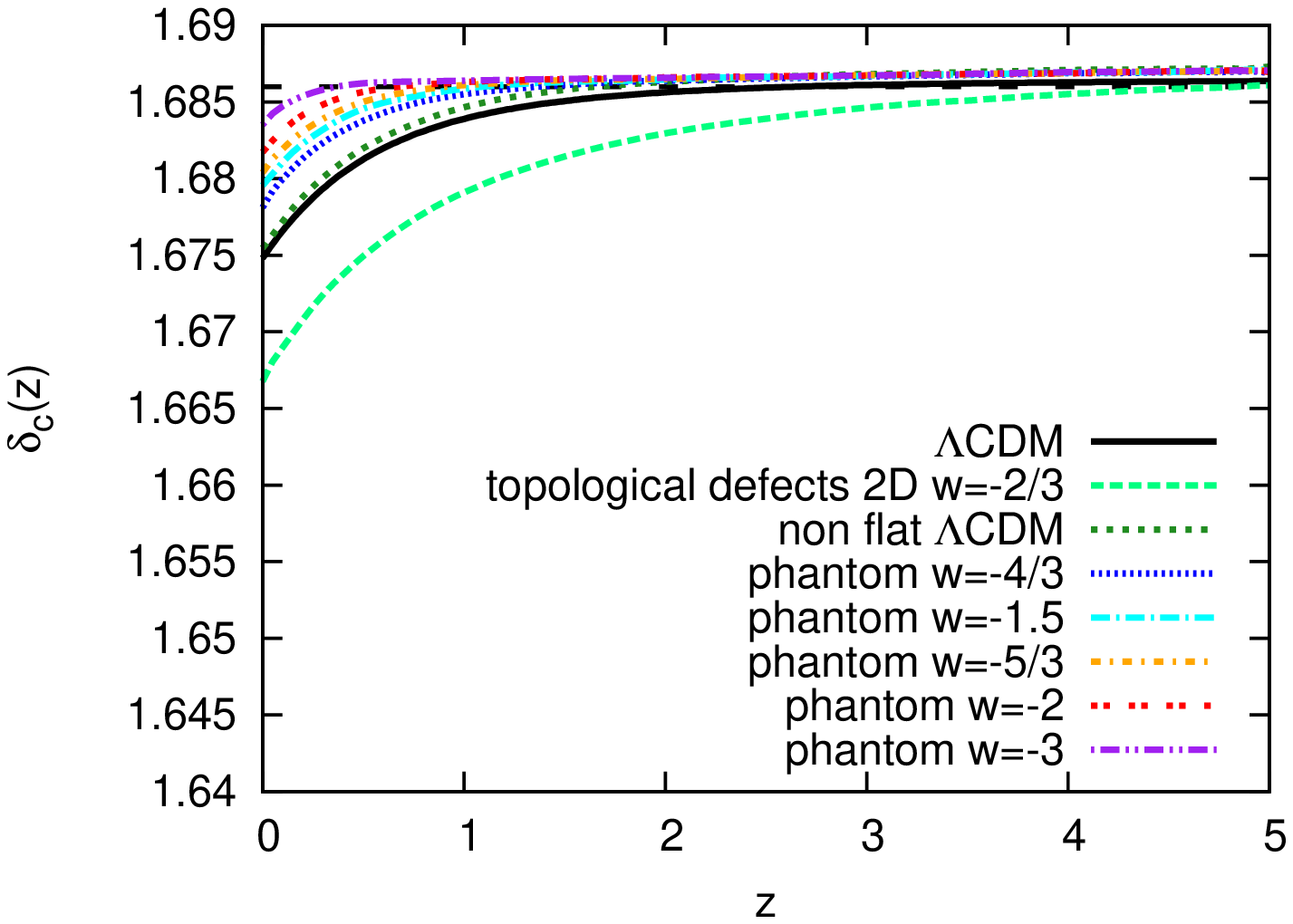}
\includegraphics[angle=-90,width=0.49\hsize]{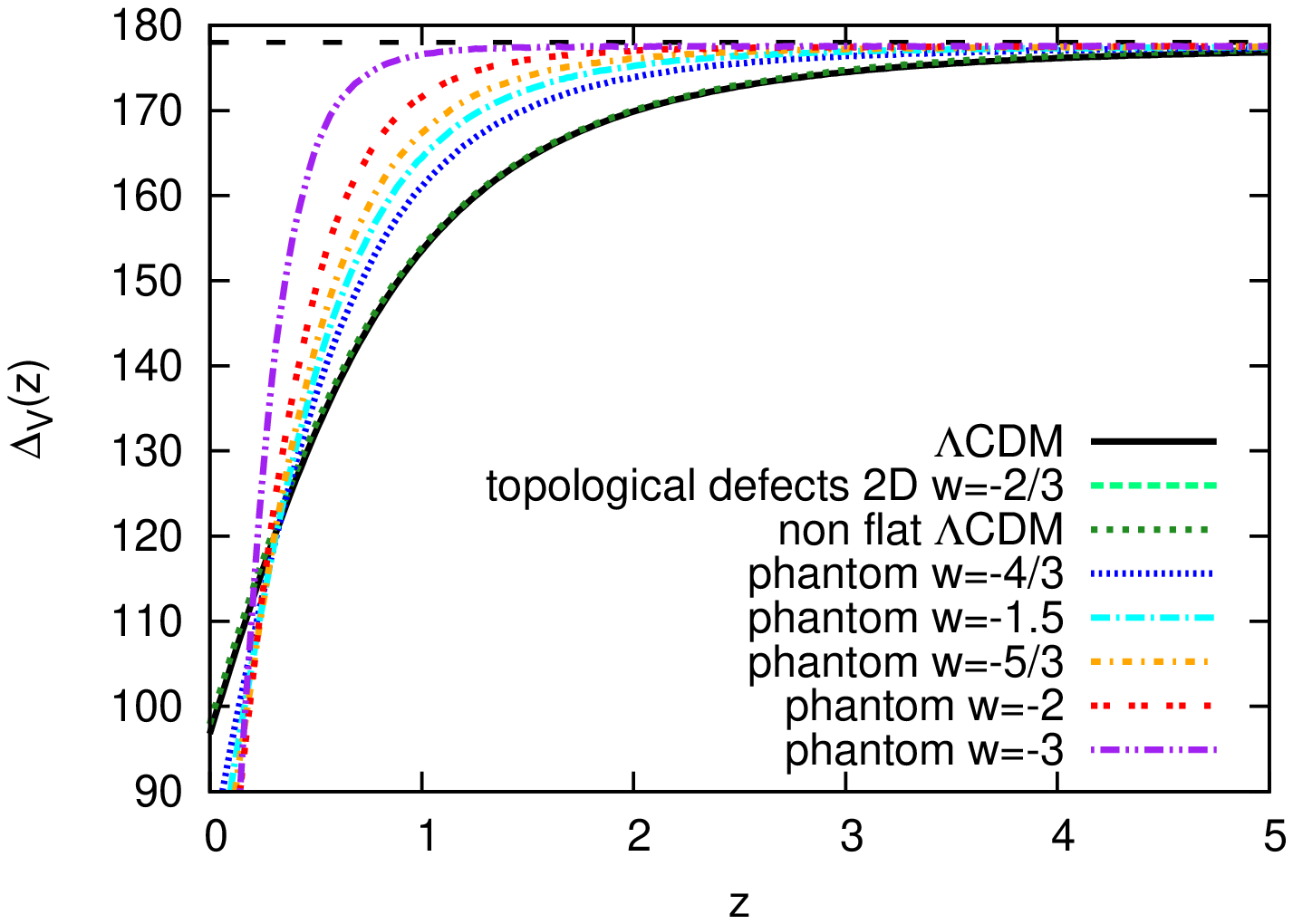}
\caption{The left panels show the time evolution of the linear overdensity $\delta_{\mathrm{c}}(z)$, the right panels the time evolution for the virial overdensity $\Delta_{\mathrm{V}}(z)$ for the different classes of models. In all panels, the $\Lambda$CDM solution (black solid curve) is the reference model, while the black dashed horizontal curve is the EdS model that is reached asymptotically by all the models. The upper panels present the quintessence models: the INV1 (INV2) model is shown with the light-green dashed (dark-green short-dashed) curve, the 2EXP model with the blue dotted curve, the AS model with the cyan dot-dashed curve, the CPL (CNR) model with the red dot-dotted (orange dot-short-dashed) curve and finally the SUGRA model with the violet dot-dot-dashed curve. The middle panels show the Casimir effect (brown dotted curve) and the (generalised) Chaplygin gas with the (turquoise short-dashed) magenta dashed curve. Finally the lower panels report the solution for the models with constant equation-of-state parameter: the dark-green short-dashed curve stands for the non-flat $\Lambda$CDM model, the light-green dashed curve for the model with $w=-2/3$, the blue dotted curve represents the model with $w=-4/3$, the cyan dot-dashed curve the model with $w=-1.5$, the orange dot-short-dashed curve the model with $w=-5/3$, the red dot-dotted curve the model with $w=-2$ and finally the violet dot-dot-dashed curve curve shows the model with $w=-3$.}
\label{fig:spc}
\end{figure*}

Our main results are shown in Fig.~\ref{fig:spc}. The right panels show results for the virial overdensity $\Delta_{\mathrm{V}}(z)$, while the left panels are specialised to the linear overdensity $\delta_{\mathrm{c}}$. The upper panels refer to the quintessence models, the middle panels refer to the (generalised) Chaplygin gas and to a cosmology with Casimir effect taken into account. The lower panels show results for the models with a constant equation-of-state parameter (non-flat $\Lambda$CDM model, topological defects and phantom models). Linestyle labels are explained in the figure caption.

From Fig.~\ref{fig:spc}, it is quite evident that all models considered, including the EDE cosmologies discussed in the appendix, behave very similarly to the flat $\Lambda$CDM cosmology, irrespective of the equation-of-state parameter, be it constant or varying with time. At $z=0$, the difference in $\delta_{\mathrm{c}}$ is at most of $2\%$ for the generalised Chaplygin gas, while for the very large majority it is even less. All models asymptotically approach the EdS limit at high redshift. This result, that all models give essentially the same results, is quite important: It shows that the linear density threshold $\delta_\mathrm{c}$ from the $\Lambda$CDM model is very close to the precise value in other cosmologies even if the equation-of-state parameter considerably differs from $w=-1$. We argue that a possible enhancement in structure formation might be caused by rapidly varying or discontinuous equation-of-state parameters, for example if they contain bumps or peaks. From a physical point of view, huge differences from the $\Lambda$CDM models might result from modified-gravity scenarios, such as coupled dark-energy models.

It is also interesting to see that the equation-of-state parameter has very little impact on the evolution of $\delta_{\mathrm{c}}$. We argue that this can be due to the fact that the equation-of-state parameter is always integrated over and thus its effects are smoothed over cosmic history. It would be interesting to work out with the equations governing the evolution of the overdensity which conditions must be satisfied by the equation-of-state parameter to have significant effects on $\delta_{\mathrm{c}}$.

The same considerations apply to the virial overdensity $\Delta_{\mathrm{V}}$. Deviations at low redshift are at most of the order of a few percent, thus having negligible impact on non-linear structure evolution. This fact has also a practical advantage: all quantities depending on $\Delta_{\mathrm{V}}$ will be virtually unaffected if the virial overdensity of the $\Lambda$CDM model is used as an approximation.

We thus conclude that the models discussed above have no significant impact on non-linear structure formation, including the EDE models presented in the appendix. Hence, the conclusions on the thermal Sunyaev-Zel'dovich effect \citep{Sadeh2007,Waizmann2009}, lensing \citep{Fedeli2007, Fedeli2008} and clustering \citep{Fedeli2009} are based on erroneous assumptions.

\section{Volume effects on haloes number counts}

In the previous section, we studied the impact of different dark energy models on the linear overdensity threshold $\delta_{\mathrm{c}}$. A quantity closely related to observations is the mass function, representing the number of collapsed objects per unit mass and volume. Since it only depends on $\delta_{\mathrm{c}}$ and on the growth factor, no appreciable differences are expected between the models studied. An important quantity that can be derived from observations is the total number of haloes above a given mass in a complete survey volume. The minimum mass detectable in a survey is generally a function of redshift and changes with the observed wave band; moreover, it will also depend on the survey according to the instrument sensitivity. Since we do not intend to specify an individual survey here, we assume the minimum mass to be independent of redshift. An idealisation in this approach is that the catalogue of objects is considered to be complete in order to compare observations with theoretical predictions.

The cumulative number of haloes above a given mass $M_\mathrm{h}$ is
\begin{equation}
  N(>M_\mathrm{h})=\int_{M_{\mathrm{min}}}^{\infty}\mathrm{d}M
  \int_{z_1}^{z_2}\frac{\mathrm{d}n}{\mathrm{d}M\mathrm{d}V}\frac{\mathrm{d}V}{\mathrm{d}z}dz\;,
\end{equation}
where $\mathrm{d}n/\mathrm{d}M\mathrm{d}V$ represents the differential mass function and $\mathrm{d}V/\mathrm{d}z$ the volume element. Since dark energy does not only affect the growth history but also the geometry, we expect that the contribution of volume effects on observable quantities will provide more information than merely the differential mass function.

\begin{figure*}
\includegraphics[angle=-90,width=0.49\hsize]{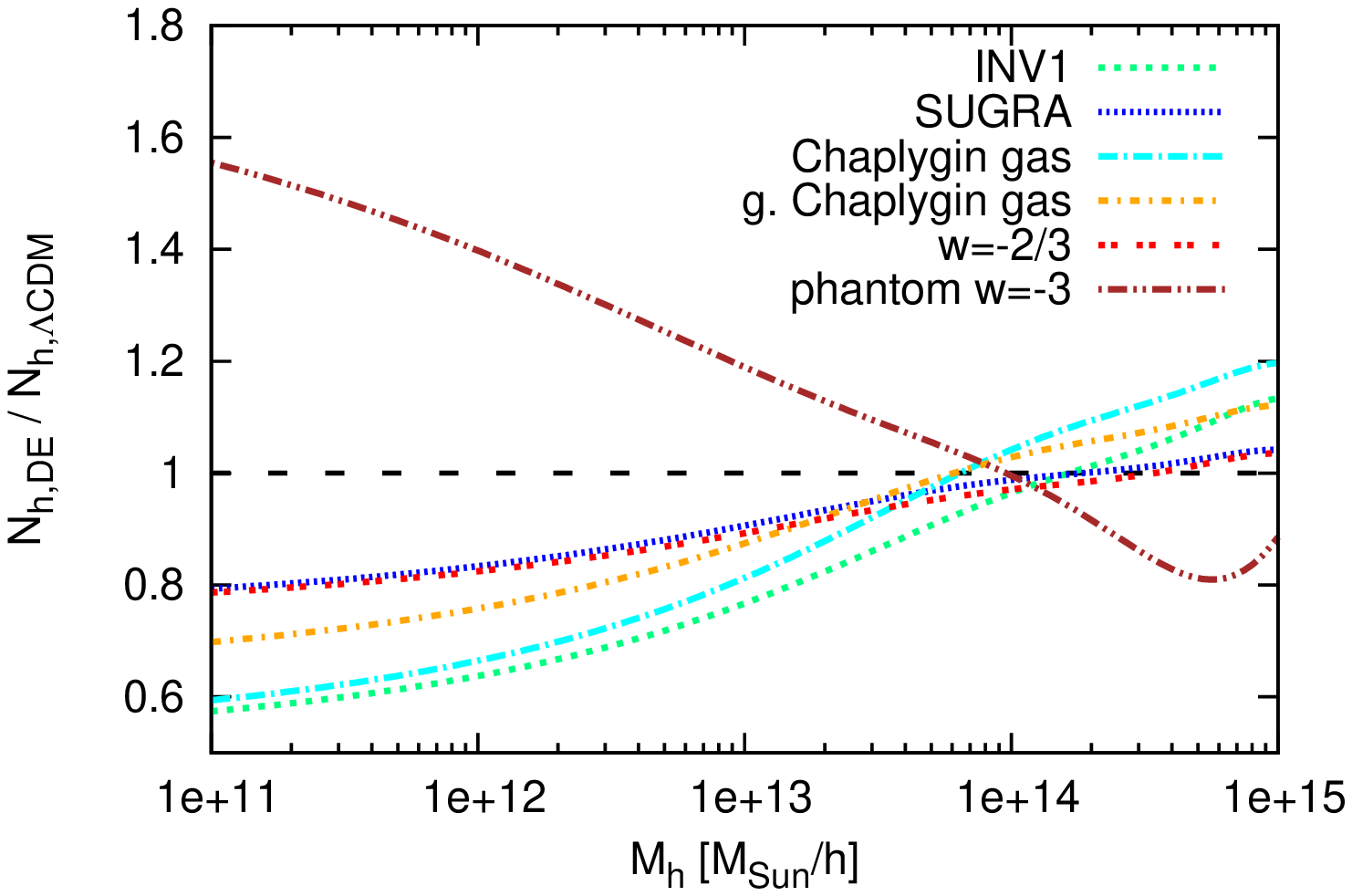}
\includegraphics[angle=-90,width=0.49\hsize]{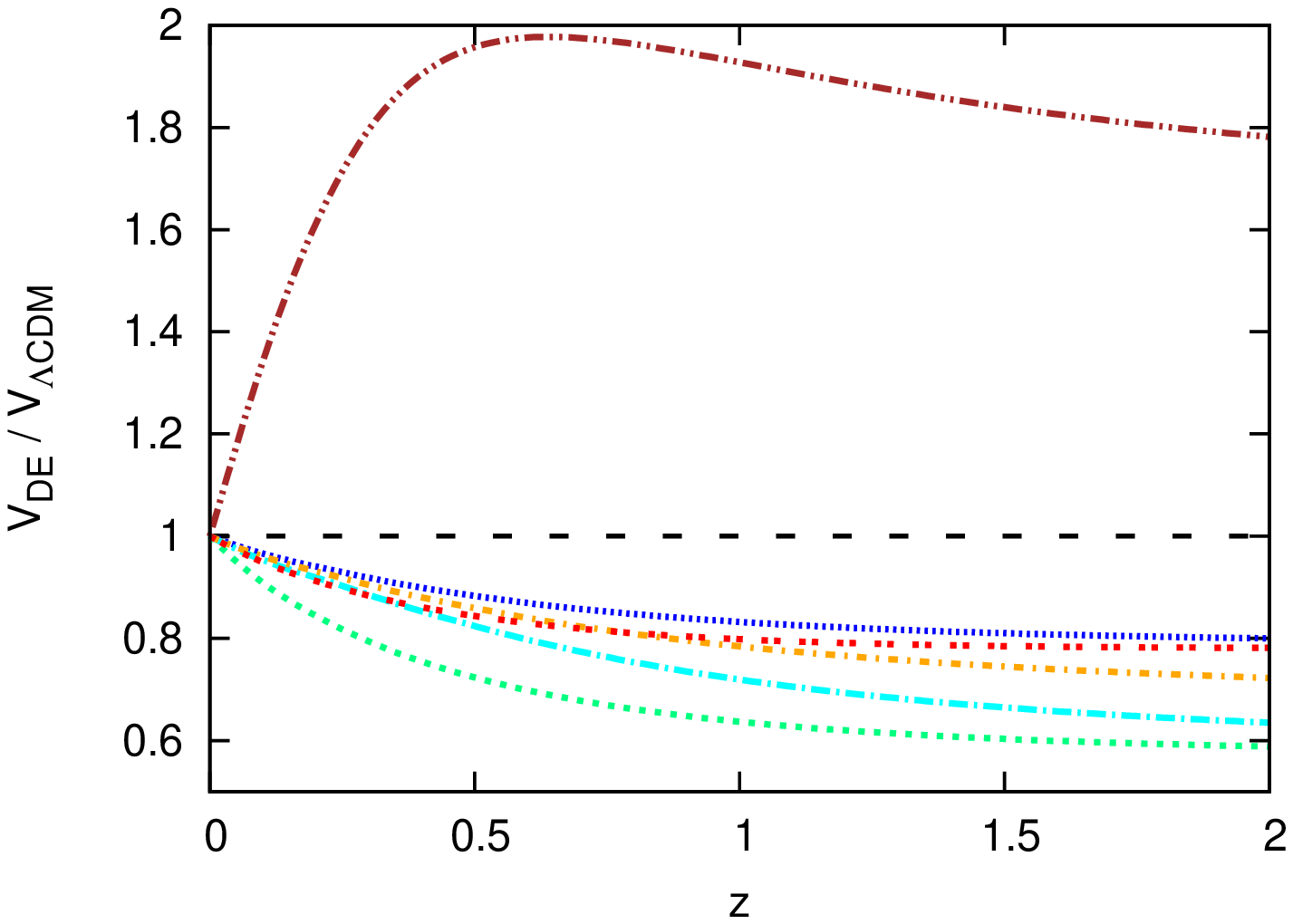}\hfill
\includegraphics[angle=-90,width=0.49\hsize]{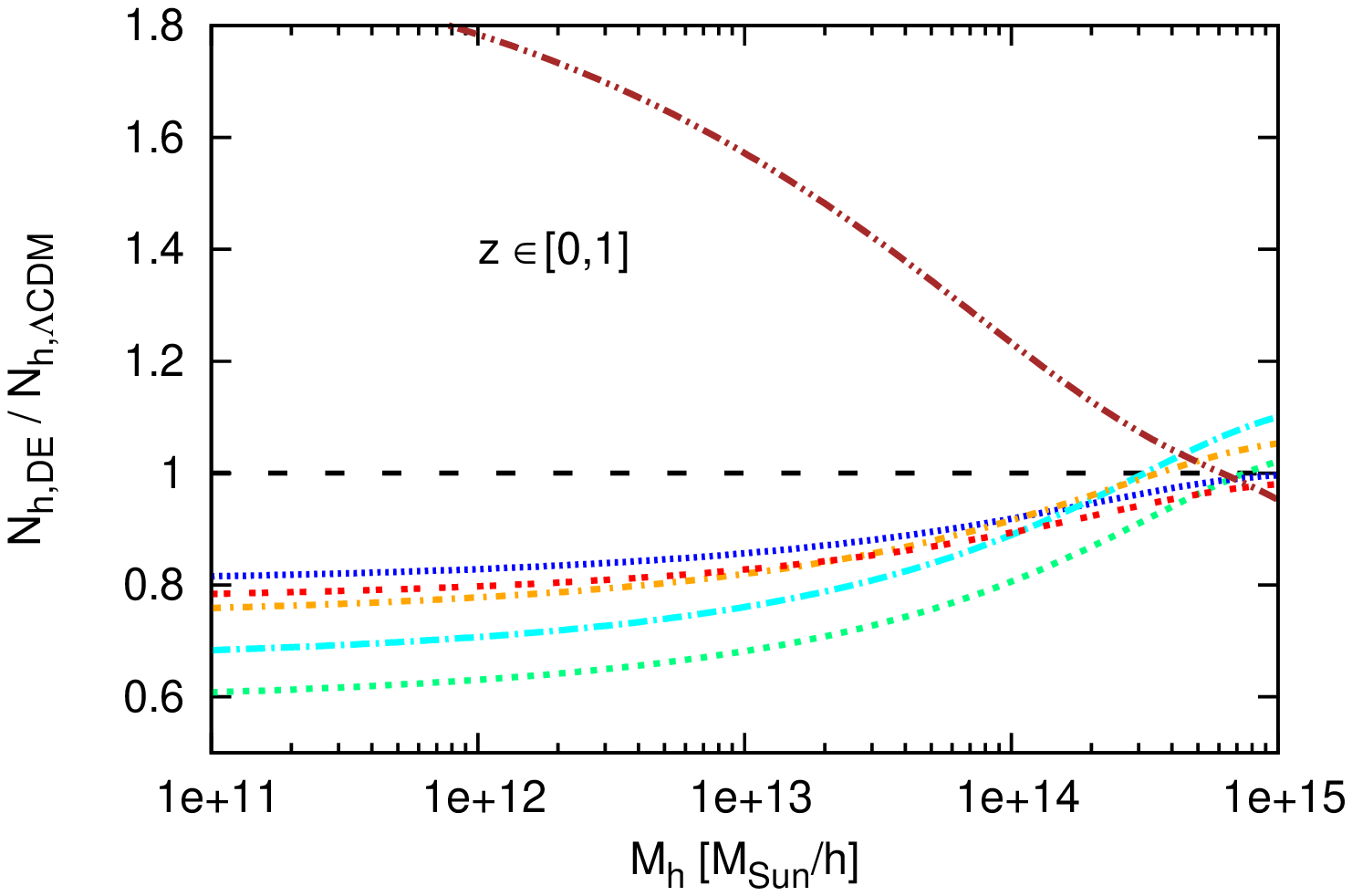}
\includegraphics[angle=-90,width=0.49\hsize]{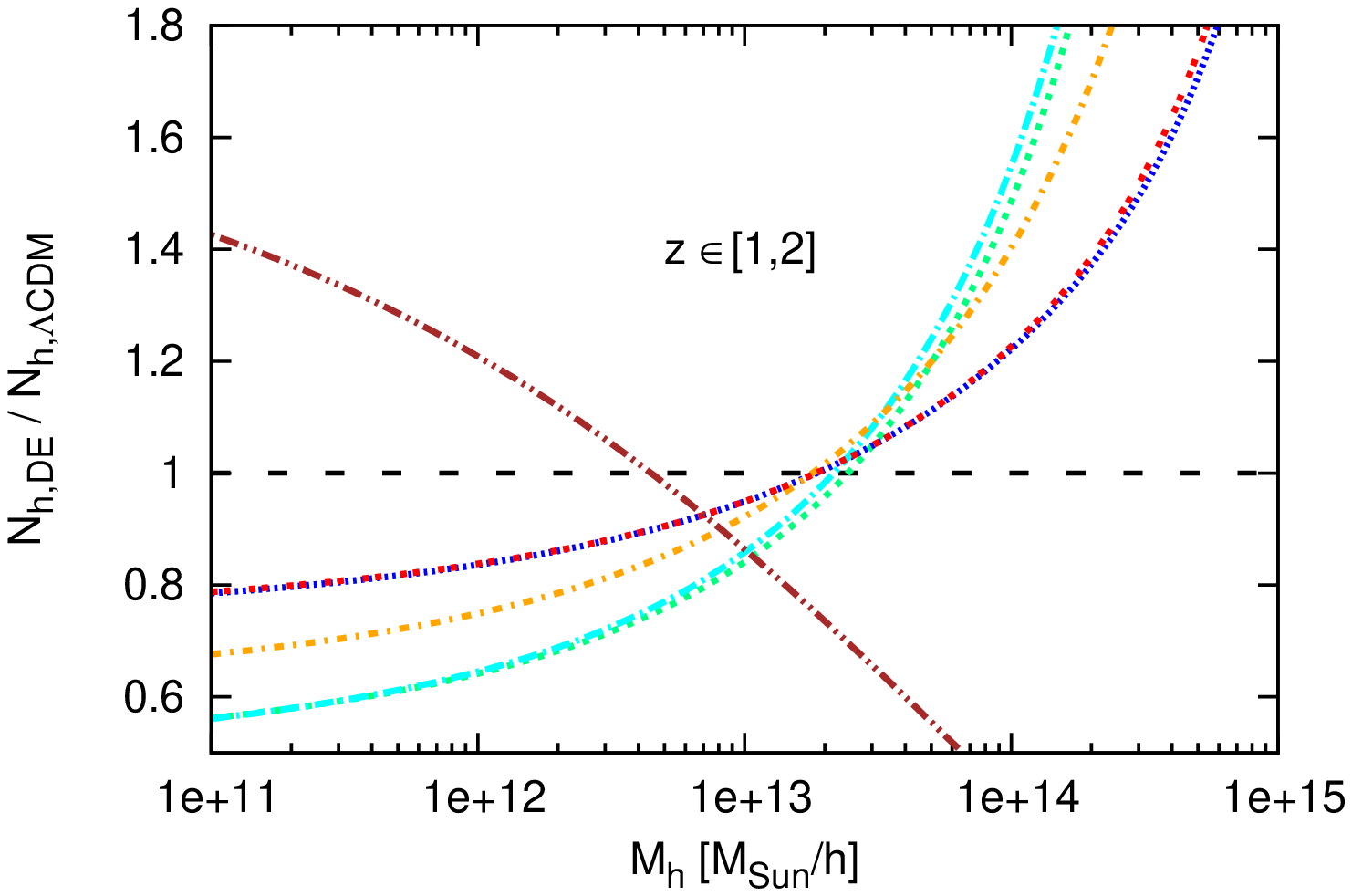}
\caption{Volume effects on halo number counts. The top left panel shows the cumulative mass function between $z=0$ and $z=2$ normalised with the expected value of the $\Lambda$CDM model. The top right panel shows the volume as a function of redshift compared to the volume of a $\Lambda$CDM model. The lower panels present the different contributions to the number counts in two redshift bins, between $z=0$ and $z=1$ and between $z=1$ and $z=2$ in the left and right panels, respectively. Different colours refer to different dark energy models. Green short-dashed: INV1; blue dotted: SUGRA; (orange dot-short-dashed) cyan dot-dashed: (generalised) Chaplygin gas; red dot-dotted (brown dot-dot-dashed): phantom model with $w=-2/3$ ($w=-3$).}
\label{fig:mf}
\end{figure*}

Effects on the number of observable haloes are shown in Fig.~\ref{fig:mf}. In the top left panel, we show the ratio of the cumulative mass function above a given mass integrated between $z=0$ and $z=2$ between some of the dark energy models studied and the reference $\Lambda$CDM model. The top right panel shows the volume effect for the corresponding models, i.e.~the ratio between the volumes of the dark-energy and the $\Lambda$CDM models. In the bottom panels, we show the contribution to the number counts in spherical shells enclosing a volume between $z=0$ and $z=1$ (left panel) and $z=1$ and $z=2$ (right panel). Please see the figure caption for details on the models considered.

In the Press \& Schechter formalism, the cosmological information, and hence the dark energy contribution, is contained in the quantity $\delta_c(z)/D(z)$, where $D(z)$ is the growth factor. To compare the different models, we thus fix the variance for the $\Lambda$CDM model ($\sigma_8=0.8$) and scale the variance of the dark energy models according to \citep[see also][]{Abramo2007}
\begin{equation}
  \sigma_{8,\mathrm{DE}}=
  \frac{\delta_{\mathrm{c,DE}}(z=0)}{\delta_{\mathrm{c,\Lambda CDM}}(z=0)}\sigma_8\;.
\end{equation}
Due to the relatively small differences in terms of $\delta_c$, the normalisations differ by a few percent at most.

Even if the differential mass functions differ only slightly, we note that differences in the number counts are as large as $40\%-60\%$. This is mainly due to volume effects, as shown in the top right panel of Fig.~\ref{fig:mf}. Models with a non-phantom equation of state always have a smaller volume than the $\Lambda$CDM model because the expansion rate is lower, while the opposite holds for phantom models (dot-dot-dashed brown curve). Since differences in the mass function are expected only in the high-mass tail, we can safely assume that the number of small objects is approximately the same for all the models. Thus, for low mass haloes, the non-phantom dark energy models predict fewer objects, but for objects above $\approx 10^{14}~M_{\odot}/h$, the exponential tail of the mass function compensates the smaller volume, and we see that a larger number of high-mass objects is expected. Of course, for the phantom models, the results are reversed. We expect more objects at low mass and fewer at high mass, since they do not have time to assemble.

From an observational point of view, it is also interesting to determine in which redshift interval we expect the highest contribution. This is shown in the lower panels. We normalise to the $\Lambda$CDM counts integrated over the same redshift interval as the dark-energy models, thus the sum of the two panels does not reproduce the top left panel. It is clear from the bottom right panel that the major contribution comes from high redshifts, while we do not expect more than $10\%$ difference from the volume up to $z=1$. Once again, for the phantom models, the situation is reversed.

Despite the fact that differences in number counts are not negligible and systematic, we have to recall that they are of the same order of magnitude as the uncertainty in the determination of halo masses. It will therefore still be difficult to discriminate between the models studied.

\section{Conclusions}\label{sect:conc}

In this work, we have generalised the non-linear equation governing the evolution of matter overdensities to cosmological models containing fluids with an arbitrary equation-of-state parameter. Specifying $w=0$, we recover the well-known equation for structure formation in matter-dominated universes. By means of the non-linear evolution equation, we determine the appropriate initial conditions used to solve the linear equation and compute the linear threshold for collapse, $\delta_{\mathrm{c}}$. We point out that the derivation of Eq.~(\ref{eqn:nldeq}) is very general and can be extended to very broad classes of cosmological models, once the appropriate continuity, Euler, Poisson and background equations are provided.

In this work, we considered exclusively non-clustering dark energy models, in which the only clustering component is the dark matter. Our goal was to study a whole catalogue of dynamical dark-energy models, thereby summarising results partly obtained elsewhere, and to clarify discrepant results on early dark-energy models. We stuck to the common assumption that the sound speed in the dark energy is given by the equation of state. There are more general scenarios allowing perturbations also in the dark-energy component, and this might lead to important differences compared to the case of non-clustering dark energy \citep{Abramo2007}. If the dark-energy component is not homogeneous, one might wonder what happens to the number counts if the equation of state of the dark-energy fluid changes within the collapsing sphere. Then, the equation of state will depend on the actual overdensity of the dark energy, and the model will acquire an additional degree of freedom, parametrised by the effective sound speed defined by $c_{\mathrm{s,eff}}^2=\delta P/\delta\rho$. Theoretical predictions for the equation of state inside the collapsing sphere are given by \cite{Abramo2008}, and a determination of the effective sound speed in Sunyaev-Zel'dovich and weak-lensing surveys is carried out in \cite{Abramo2009.1}. The authors found that a negative pressure perturbation to the dark energy fluid may have a substantial effect on the number counts as shown in their Fig.~1. A negative effective sound speed $c_s^2$ can lower the value of $\delta_\mathrm{c}$ to $\approx1.5\ldots1.55$, giving a substantial boost to structure formation. Volume effects will still be comparable with what we found here. It may then be possible to descriminate this class of models compared to those studied here.

Despite the fact that the differential mass functions are very similar because of the small differences in $\delta_{\mathrm{c}}$, we still expect a significant difference for the total number of objects in a given volume above a given mass threshold. For low minimum mass, the volume effects dominate and we expect a lower overall number of objects, while the mass function dominates over the volume effect at the high-mass tail, and we still expect more haloes. Unfortunately, as shown in the lower panels of Fig.~\ref{fig:mf}, the major contributions come from redshifts above $z=1$, where fewer objects are expected and where observations are more difficult.

We found that all models studied here show differences at the per-cent level compared to the standard $\Lambda$CDM model. We also argue that in the framework of general relativity, it may be possible to have a more pronounced impact on non-linear structure formation if the equation-of-state parameter is discontinuous or if modifications to general relativity are involved. This will be the subject of future work.

We thus conclude that, at least for the wide class of models studied here and due to the current status of observations, it is quite difficult to use number counts to discriminate between different dark-energy models. A discrimination may still be possible to eventually based on geometrical tests where volume effects are relevant.

\section*{Acknowledgements}

We are grateful to Christoph Wetterich and Valeria Pettorino for useful discussions and for suggesting the approach based on Newtonian hydrodynamics, and to Nelson Nunes and Andrea Macci{\`o} for useful discussions and comments on the manuscript. We also thank the anonymous referee whose comments helped to improve the paper. Francesco Pace is also indebted to Margherita Grossi for providing the data of her numerical simulations. This work was supported by the Deutsche Forschungsgemeinschaft (DFG) under the grants BA 1369/5-1 and 1369/5-2 and through the Transregio-Sonderforschungsbereich TR 33, as well as by the DAAD and CRUI through their Vigoni programme.

\bibliographystyle{mn2e}

\appendix

\section{$\delta_{\mathrm{c}}$ for EDE cosmologies}\label{sect:app}

Having found that cosmological models with dynamical dark energy have very little impact on the spherical-collapse parameters $\delta_\mathrm{c}$ and $\Delta_\mathrm{V}$, we have to clarify and explain how we could arrive at contradictory results in \cite{Bartelmann2006}.

We first show the results we obtain with the new approach presented in this paper for the linear overdensity $\delta_{\mathrm{c}}$ in two different EDE models. We further compare our theoretical results to the numerically simulated mass function of \cite{Grossi2009} before we turn to explain why the earlier calculation of $\delta_{\mathrm{c}}$ \citep{Bartelmann2006} arrived at significantly different results.

\begin{figure}
\includegraphics[angle=-90,width=0.9\hsize]{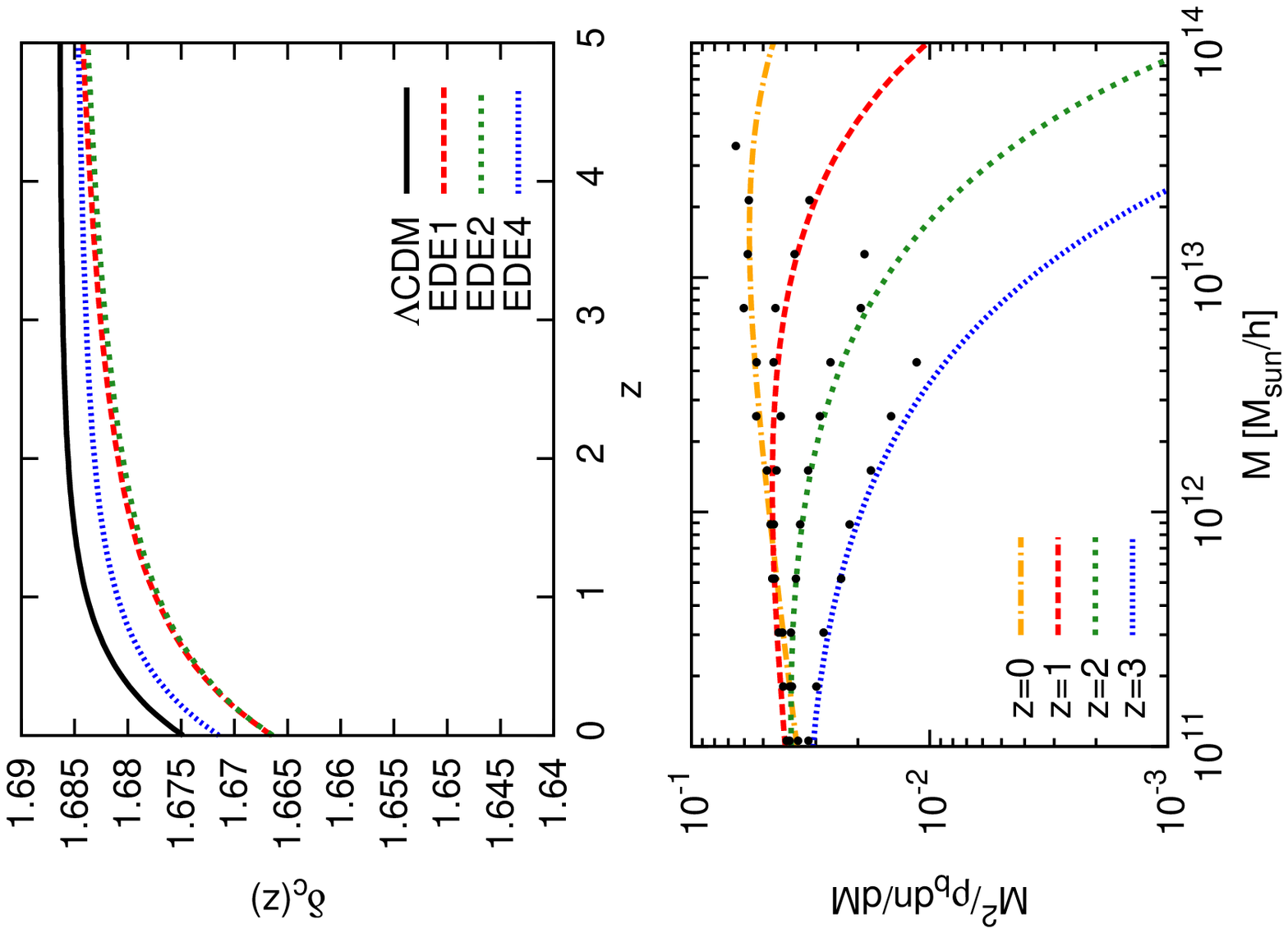}
\caption{Upper panel: $\delta_c$ for three different early dark energy models. The black solid curve represents the $\Lambda$CDM model while the red dashed, green short-dashed and blue dotted curves represent the EDE1, EDE2 and EDE4 models, respectively. Lower panel: comparison between the theoretical multiplicity mass function (given by the differential mass function times the mass squared) and the numerical one. Results for four different redshifts are shown: $z=0$ (orange dot-dashed), $z=1$ (red dashed), $z=2$ (green short-dashed) and $z=3$ (blue dotted). Points represent the data of the $N$-body simulation.}
\label{fig:A1}
\end{figure}

Figure~\ref{fig:A1} shows the time evolution of $\delta_{\mathrm{c}}$ as a function of redshift (upper panel) as well as the comparison between the theoretical prediction for the mass function using the Sheth-Tormen expression \citep{Sheth1999} and the numerical mass functions obtained via N-body simulations at the redshifts $z\in\{0, 1, 2, 3\}$ (lower panel).

We notice that EDE cosmologies do not have any strong impact on the spherical-collapse parameters and neither on the halo mass function. The deviations from $\Lambda$CDM reach at most of $(1\ldots2)\%$, in excellent agreement with numerical simulations.

In the earlier study by \cite{Bartelmann2006}, several approximations had to be made to render the spherical-collapse equations numerically tractable and stable. In contrast to the new approach presented here, the main problem there was that the spherical-collapse equations become singular at times $t\to0$, while $\delta_\mathrm{c}$ must be obtained extrapolating from the limit of the solution for $t\to0$. Yet, the solutions for the radius $y(x)$ of the spherical overdensity as a function of the scale factor $x$ as well as for the density contrast $\delta(x)$, turn out to agree precisely with the solutions obtained with the new approach.

\begin{figure}
\includegraphics[width=\hsize]{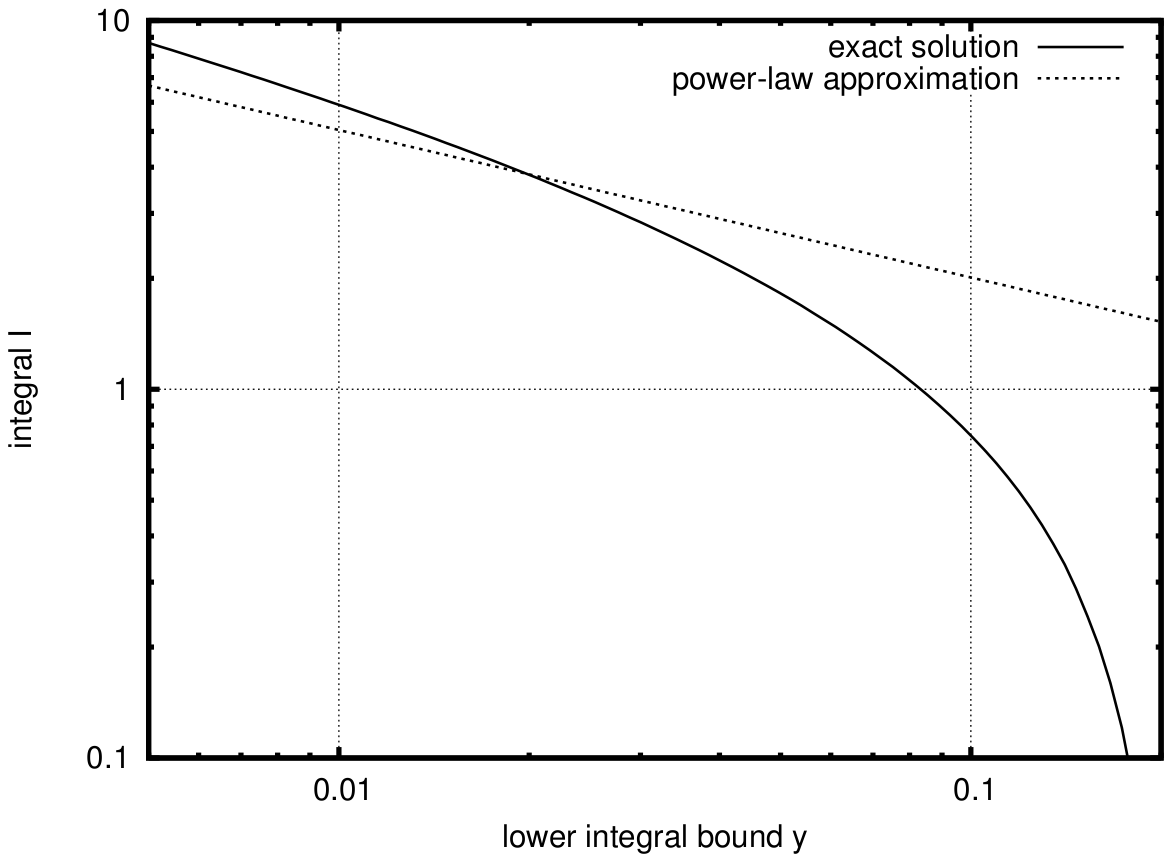}
\caption{Exact solution of the integral $I$ from Eq.~(15) in \protect\cite{Bartelmann2006} compared to the power-law approximation defined in Eq.~(19) there. The failure of this approximation is the reason for the discrepant results obtained there compared to the present study.}
\label{fig:A2}
\end{figure}

Substantial deviations begin with the integration constant $B$ introduced in Eqs.~(22) and (23) in \cite{Bartelmann2006}. According to the approximations used there, $B\approx1$ to good accuracy. In the new approach, this can be tested using $\delta=3By/5$ at early times. It turns out that the correct result is $B\approx2.13$. This substantial deviation can be traced back to the integrand $(1+3w)g(x)y$ of the integral $I$ in Eq.~(15). The power-law approximation made in Eq.~(18) turns out to hold very well for early times, but to fail considerably at late times. Thus, the late-time evolution of the integrand in Eq.~(15) is incorrect, and the deviation from its exact behaviour starts becoming substantial already at scale factors $x\gtrsim0.1$. The power-law approximation of Eq.~(19) for the integral $I$ is accordingly incorrect. The assumed power-law behaviour and its true shape are shown in Fig.~\ref{fig:A2}.

Intriguingly, the approximation is correct for conventional Friedmann models with arbitrary cosmological constant and also for cosmologies with dynamical dark energy with vanishing dark-energy density at early times. Its inaccuracy thus remained undiscovered in the numerous tests that were carried out.

\label{lastpage}

\end{document}